\documentclass[twocolumn]{aa}
\usepackage{graphicx}
\usepackage{txfonts}
\begin{document}
   \title{Multi-colour optical monitoring of eight red blazars
   \thanks{Table 2 is only available in electronic at the CDS via
anonymous ftp to cdsarc.u-strasbg.fr (130.79.125.5) or via
http://cdsweb.u-strasbg.fr/Abstract.html}}

%Table 2 is only available in electronic form at the
%    CDS via anonymous ftp to cdsarc.u-strasbg.fr (130.79.128.5)

   %\subtitle{I. Overviewing the $\kappa$-mechanism}

   \author{Minfeng Gu
          \inst{1,2,3}
          \and C.-U. Lee\inst{1}
          \and Soojong Pak\inst{1}
          \and H. S. Yim\inst{1}%\fnmsep\thanks{Just to show the usage
%          of the elements in the author field}
          \and A. B. Fletcher\inst{1,2}
          }

   \offprints{M. F. Gu, e-mail: gumf@shao.ac.cn}

   \institute{Korea Astronomy and Space Science Institute, 61-1 Whaam-dong,
   Yuseong-gu, Daejeon 305-348, Republic of Korea
%              \\              \email{wuchterl@amok.ast.univie.ac.at}
         \and Shanghai Astronomical Observatory, Chinese Academy
         of Sciences, 80 Nandan Road, Shanghai 200030, China
%             \\             \email{c.ptolemy@hipparch.uheaven.space}
%             \thanks{The university of heaven temporarily does not
%                     accept e-mails}
         \and National Astronomical Observatories, Chinese Academy
         of Sciences, Beijing 100012, China
             }

   \date{Received; accepted;}

   \abstract{We present the observational results of multi-colour optical
   monitoring of eight
   red blazars from 2003 September to 2004 February.}
   {The aim of
   our monitoring is to investigate the spectral variability as well as
   the flux variations at short and long time scales.}
   {The observations were carried out using the 1.0 m
   robotic telescope of Mt. Lemmon Optical Astronomy Observatory, in Arizona,
   USA, the 0.6 m telescope of Sobaeksan Optical Astronomy Observatory and
   the 1.8 m telescope of Bohyunsan Optical Astronomy Observatory, in the
   Republic of Korea.}
   {During the observations, all sources show strong flux variations with amplitudes of larger than 0.5 mag.
   Variations with amplitudes of over 1 mag are found in four sources. Intraday
   variations with amplitudes larger than 0.15 mag, and a rapid brightness increase
   with a rate of $\sim0.2$ mag per day in four days, are
   detected in S5 0716+71. We investigate the relationship between the colour
   index and source brightness for each source. We find that two out of three
    FSRQs tend to be redder when they are brighter, and, conversely, all BL
    Lac objects tend to be bluer. In particular, we
   find a significant anti-correlation between the $V-I$ colour index and
   $R$ magnitude for 3C 454.3.
   This implies that the spectrum became steeper when the source was
   brighter, which is opposite to the common trend for blazars. In contrast,
   significant positive
   correlations are found in 3C 66A, S5 0716+71, and BL Lac. However, there are
   only very weak correlations for PKS 0735+17 and OJ
   287.}
   {We propose that the different relative contributions of the
   thermal versus non-thermal radiation to the optical emission
   may be responsible for the different trends of the colour
   index with brightness in FSRQs and BL Lac objects.}
   %, at least for red blazars.

   %We propose that
   %the contribution of the thermal emission from the accretion
   %disk in the optical region may be responsible for the different dependence of the
   %colour indices on the source brightness in FSRQs and BL Lac
   %objects.


   \maketitle
%
%________________________________________________________________

\section{Introduction}

Blazars, including BL Lac objects and flat-spectrum radio quasars
(FSRQs), are the most extreme class of active galactic nuclei
(AGNs), characterized by strong and rapid variability, high
polarization, and apparent superluminal motion. These extreme
properties are generally interpreted as a consequence of
non-thermal emission from a relativistic jet oriented close to the
line of sight. As such, they represent a fortuitous natural
laboratory with which to study the physical properties of jets,
and, ultimately, the mechanisms of energy extraction from the
central supermassive black holes.
%, a fundamental goal of extragalactic astrophysics.

In the field of AGN observations, one of the most important
discoveries in the last decade has been that blazars emit a
substantial fraction, sometimes most of their power, at
$\gamma$-ray energies (GeV and TeV, e.g. Ulrich, Maraschi \& Urry
1997; Catanese \& Weeks 1999). The $\gamma$-ray emission of
blazars indicates a double-peak structure in the overall spectral
energy distribution (SED), with two broad spectral components. The
first, lower frequency component is generally interpreted as being
due to synchrotron emission, and the second, higher frequency one
as being due to inverse Compton emission. Although different
blazars have different peak frequencies, the two peaks in their
SEDs are separated by approximately the same amount, i.e. 8$-$10
decades in frequency (Fossati et al. 1998). According to the
different peak frequencies, blazars are divided into two
subclasses: the low-energy-peaked blazars (red blazars), which
have synchrotron peaks in the IR/optical range, and the
high-energy-peaked blazars (blue blazars), which have synchrotron
peaks at UV/X-ray energies. All of the TeV $\gamma$-ray-loud AGNs
are blue blazars, while all strong GeV $\gamma$-ray AGNs are red
blazars (Mattox et al. 1997; Hartman et al. 1999; Bai \& Lee
2001).

Although a point of agreement about the second component is that
the $\gamma$-rays of blazars are produced in relativistic jets by
inverse Compton scattering, the origin of the seed photons
(optical/IR), the location and size of the emitting region, and
the degree of relativistic beaming of the high-energy radiation,
are all still unknown. Seed photons may come from the jet itself
(Synchrotron Self-Compton model), from an accretion disk around a
supermassive black hole at the base of the jet, or else from
photons of the broad emission line region (e.g. Wehrle et al.
1998; Collmar et al. 2000). Some models suggest that the flares of
$\gamma$-rays in blazars are caused by the increase and decrease
of the bulk Lorentz factor, while other models suggest changes in
both the injection rate and the spectral shape of the injected
electrons (Mukherjee et al. 1997; Hartman et al. 2001; Pian et al.
2002; Villata et al. 2002).

Simultaneous multiwavelength observations have shown that both TeV
and GeV $\gamma$-ray blazars vary much more violently at and above
their peaks, than below these peaks, and that flux variations at
these two peaks are correlated, indicating that both components
are produced by the same electron population in the relativistic
jet (e.g. Ulrich et al. 1997; Sambruna et al. 2000; Hartman et al.
2001). Since red blazars have synchrotron emission peaking in the
IR/optical bands, observations at these wavelengths are thus
crucial in understanding the nature of $\gamma$-ray emission in
red blazars. The near-IR/optical bands are ideal for monitoring
the flux and spectral variability, as well as to determine the
frequencies and frequency shifts of the synchrotron peaks for red
blazars. The flux and spectral variability, together with the
synchrotron peak frequency and its shift, allow one to derive
information on the radiating relativistic electrons, and on the
emission region (Ghisellini et al. 1997). Moreover, the
differences in the variability behaviour, in the IR/optical bands,
of non-$\gamma$-ray-detected red blazars, as compared to those of
the $\gamma$-ray-loud ones, may strongly constrain models for
$\gamma$-ray emission.
%and can be used to select the $\gamma$-ray
%source candidates which may be observed by $\gamma$-ray
%observatories, such as the GLAST, in the near future. The strong
%TeV $\gamma$-ray sources, Mrk 501, Mrk 421, and 1ES 2344+514, and
%the strong GeV $\gamma$-ray source, 3C 279, have exhibited large
%spectral variability during $\gamma$-ray flares, while OJ 287,
%which was only marginally detected by EGRET even during the recent
%large outburst, showed rather stable spectra during this outburst.
%Spectral variability and in situ particle acceleration are likely
%to be necessary conditions for a blazar to be a strong
%$\gamma$-ray source. In addition, as is well known, the flux
%variability in the IR/optical bands provides constraints on the
%size of the $\gamma$-ray emission region.

The study of the properties of the SED is an excellent diagnostic
tool for theoretical models. However, the availability of
simultaneous multiwavelength observations is generally inadequate
in tracing the time evolution for both the synchrotron and inverse
Compton components. A relatively simple method is to investigate
time evolution in a selected region of the spectra, such as in the
optical. Although the optical region is narrow with respect to
other spectral regions, it may provide a relatively large amount
of information. In this case, the possible presence of other
components in addition to the synchrotron emission, e.g. thermal
emission from the accretion disk around the central engine, may
complicate the situation. Nevertheless, statistical analysis of
the optical spectral variability could enable us to constrain
theoretical models; as claimed by Vagnetti, Trevese \& Nesci
(2003), the spectral variability, even when restricted to the
optical band, can be used to set limits on the relative
contribution of the synchrotron component and the thermal
component to the overall SED.

%Although red blazars have been observed for decades, little is
%still known about the spectral variability and the shifts of the
%synchrotron peak in red blazars, especially for FSRQs, most of
%which have their synchrotron emission peaking in the IR bands.
%, let
%alone the comparison between non-gamma-ray-detected and
%gamma-ray-loud red blazars.
In this paper, we investigate the spectral variability, as well as
the multi-band flux variability for eight typical red blazars,
which consist of three FSRQs and five BL Lac objects. The spectral
variability of the FSRQs is compared with that of the BL Lac
objects. Based on analysis of the spectral variability, we attempt
to derive information to constrain the theoretical models. The
observations and data reduction are given in Sect. 2. The flux
variability is shown in Sect. 3, and the spectral variability is
described in Sect. 4. The discussions are given in Sect. 5, and
the main results are summarized in Sect. 6.

%__________________________________________________________________

\section{Observations and data reduction}

Multi-site photometric observations of eight blazars were carried
out for a combined total of 50 nights, from 2003 September to 2004
February, using the 1.0 m robotic telescope of Mt. Lemmon Optical
Astronomy Observatory (denoted as L in Table 1; Han et al. 2005),
in Arizona, USA, for 33 nights, the 0.6 m telescope of Sobaeksan
Optical Astronomy Observatory (denoted as S in Table 1) and the
1.8 m telescope of Bohyunsan Optical Astronomy Observatory
(denoted as B in Table 1), in the Republic of Korea, for 14 and 3
nights, respectively. The observations were made in 5 sessions to
investigate both long-term and intraday variations: 5 nights in
September, 10 nights in October, 13 nights in November, 16 nights
in December, 2003, and 6 nights in February, 2004. All telescopes
were equipped with CCD cameras, and ${\it}UBV$ (Johnson) and
${\it}RI$ (Cousins) filter sets. LN2 cryogenic systems were used
for both the 1.8 m and 0.6 m telescopes, while a thermoelectric
cooling system was used for the 1.0 m telescope. These three
telescopes cover different field of views: 22.2$\times$22.2
arcmin, 20.5$\times$20.5 arcmin, and 11.7$\times$11.7 arcmin for
the 1.0 m, 0.6 m, and 1.8 m telescopes, respectively. Considering
the aperture sizes of the telescopes and the seeing conditions,
different exposure times were applied. While observing, we tried
to position each target object at the same location on the CCD
surface, to within a few pixels, in order to achieve more
efficient and accurate photometry.
%While
%observing, we tried to locate the same observing region on the
%same CCD surface within few pixels to perform precise photometry.
%and observed Landolt (1992) standard star fields to correct
%systematic errors between observatories.
Sky flat images were taken at both dusk and dawn when available.
Bias and dark images were taken after obtaining sky flat images.

All images were preprocessed with the standard procedures, using
the IRAF\footnote{IRAF is distributed by the National Optical
Astronomy Observatory, which is operated by the Association of
Universities for Research in Astronomy, Inc., under cooperative
agreement with the National Science Foundation.} software,
including bias and dark subtraction, and flat-fielding. For images
obtained from the 0.6 m telescope, the fringe pattern was
carefully corrected. We used the APPHOT package in IRAF to measure
the instrumental magnitudes of the blazars and comparison stars.
The seeing variations during one night may cause variations of the
instrumental magnitude, and there is a trend of the instrumental
magnitude variations with the FWHM variations of a star from night
to night (Clements \& Carini 2001). In order to correct the seeing
variation effect during one night, and to try to include the
`total' flux from objects when seeing varies (especially when
seeing is poor), we set the radius in the aperture photometry to
be proportional to the FWHM of bright, well-exposed stars in the
image frames containing the objects, during one night (Lee et al.
2003). After experimenting with various aperture sizes, we set the
aperture radius at 4$\times$FWHM, the inner radius of the sky
annulus at 5$\times$FWHM, and the width of the sky annulus at 10
pixels, to maximize the S/N ratio.
%In order to correct the systematic differences of magnitude and color
%between observatories, we determined coefficients and applied them
%to each observatory.

In Table 1, the source IAU name and the alias name are given in
Cols. 1 and 2, respectively. The classification, filter sets,
telescopes, and the detection in the $\gamma$-region for each
source are shown in Cols. 3-6. The source magnitude is calibrated
with respect to the standard stars, for which references are given
in Col. 7 of Table 1. The observational errors are estimated from
the rms differential magnitude between the calibration star and
another standard star used for checking,
\begin{equation}
   \sigma = \sqrt{\frac{\sum(m_{i}-\overline{m})^2}{N-1}} \,,
%      \tau_{\mathrm{co}} = \frac{E_{\mathrm{th}}}{L_{r0}} \,,
   \end{equation}
where $m_{i}=(m_{C}-m_{K})_{i}$ is the differential magnitude of
the calibration star C and check star K, while
$\overline{m}=\overline{m_{C}-m_{K}}$ is the differential
magnitude averaged over the entire data set, and $N$ is the number
of observations on a given night. For each object, the typical rms
error is between 0.01 and 0.02 mag.

\begin{table*}
\caption{List of the eight blazar sources in the optical
monitoring sample. }
\begin{tabular}{lcccccc}
\hline \hline
Object & Name & Class. & Filters & Observatory & $\gamma$-detected & Comparison Stars\\
\hline
            0219$+$428 & 3C 66A         & BL Lac & $BVRI$ & L,S   & Yes & Fiorucci \& Tosti $(1996)^{\mathrm{b}}$ \\
            0420$-$014 & PKS 0420$-$01 & FSRQ   & $VRI $ & L,S   & Yes & Smith \& Balonek (1998) \\
            0716$+$714 & S5 0716+71    & BL Lac & $BVRI$ & B,L,S & Yes & Villata et al. $(1998\rm a)^{\mathrm{a}}$ \\
            0735$+$178 & PKS 0735+17   & BL Lac & $VRI $ & L,S   & Yes & Smith et al. (1985) \\
            0851$+$202 & OJ 287         & BL Lac & $VRI $ & L,S   & Yes & Fiorucci \& Tosti (1996) \\
            1641$+$399 & 3C 345         & FSRQ   & $VRI $ & L,S   & No  & Smith et al. $(1985)^{\mathrm{b}}$ \\
            2200$+$420 & BL Lac         & BL Lac & $VRI $ & L,S   & Yes & Fiorucci \& Tosti $(1996)^{\mathrm{b}}$ \\
            2251$+$158 & 3C 454.3       & FSRQ   & $VRI $ & L,S   & Yes & Fiorucci et al. (1998) \\

%            0528$+$134 & PKS 0528+134   & BVRI  & L    & &  \\
\hline
%            0235$+$164 & AO 0235+164    & BVRI & L     & Yes &  \\
\end{tabular}

\begin{list}{}{}
\item[$^{\mathrm{a}}$] $I$-data from Ghisellini et al. (1997)
\item[$^{\mathrm{b}}$] Magnitudes from Gonz\'{a}lez-P\'{e}rez et
al. (2001)
\end{list}

\end{table*}

\section{Flux variability}
The results of our monitoring program are listed in Table 2,
available in electronic form at the CDS. Column 1 is the object
name, Col. 2 is the Julian Date, Col. 3 is the magnitude, Col. 4
is the rms magnitude error, Col. 5 is the filter used, and Col. 6
is the observatory. For each object, the historical photometric
results are briefly presented. The overall magnitude variations of
each object are investigated. Variations on time scales of days,
and on shorter time scales (e.g. hours), in several objects are
also reported.

\begin{table*}
\caption{The observational results. }
\begin{tabular}{lcccccc}
\hline \hline
 Name & JD & Magnitude & Error & Filters & Observatory \\
\hline
%            0219$+$428 & 3C 66A         & BL Lac & $BVRI$ & L,S   & Yes & Fiorucci et al. $(1996)^{\mathrm{b}}$ \\
%            0420$-$014 & PKS 0420$-$014 & FSRQ   & $VRI $ & L,S   & Yes & Smith et al. (1998) \\
%            0716$+$714 & S5 0716+714    & BL Lac & $BVRI$ & B,L,S & Yes & Villata et al. $(1998)^{\mathrm{a}}$ \\
%            0735$+$178 & PKS 0735+178   & BL Lac & $VRI $ & L,S   & Yes & Smith et al. (1985) \\
%            0851$+$203 & OJ 287         & BL Lac & $VRI $ & L,S   & Yes & Fiorucci et al. (1996) \\
%            1641$+$395 & 3C 345         & FSRQ   & $VRI $ & L,S   & No  & Smith et al. $(1985)^{\mathrm{b}}$ \\
%            2200$+$420 & BL Lac         & BL Lac & $VRI $ & L,S   & Yes & Fiorucci et al. $(1996)^{\mathrm{b}}$ \\
%            2251$+$158 & 3C 454.3       & FSRQ   & $VRI $ & L,S   & Yes & Fiorucci et al. (1998) \\

%%            0528$+$134 & PKS 0528+134   & BVRI  & L    & &  \\
\hline
%%            0235$+$164 & AO 0235+164    & BVRI & L     & Yes &  \\
\end{tabular}

%\begin{list}{}{}
%\item[$^{\mathrm{a}}$] $I$-data from Ghisellini et al. (1997)
%\item[$^{\mathrm{b}}$] Magnitude from Gonz\'{a}lez-P\'{e}rez et
%al. (2001)
%\end{list}

\end{table*}

%\subsection{}

\subsection{3C 66A}

The blazar 3C 66A is a member of the Einstein Slew Survey sample
(Elvis et al. 1992). Both GeV (Dingus et al. 1996; Mukherjee et
al. 1997) and TeV (Neshpor et al. 1998) $\gamma$-rays have been
detected in this highly polarized (Mead et al. 1990) blazar. This
source has displayed infrared, optical, ultraviolet, and X-ray
variability on different time scales (Ghosh \& Soundararajaperumal
1995; De Diego et al. 1997; Carini, Noble, \& Miller 1998). 3C 66A
was the target of an extensive multiwavelength Whole Earth Blazar
Telescope (WEBT) monitoring campaign from July 2003 through April
2004, involving observations in radio, infrared, optical, X-ray,
and very-high-energy $\gamma$-ray bands (B\"{o}ttcher et al.
2005). At all optical frequencies, a gradual brightening of the
source over the course of the campaign was observed, with a
maximum at $R\approx13.4$ on Feb. 18, 2004. Microvariability with
flux changes of $\sim5~\%$ on time scales as short as $\sim2$ hr,
and several bright flares on time scales of several days, were
detected (B\"{o}ttcher et al. 2005). Our observational data on
this source have been included in the WEBT campaign paper
(B\"{o}ttcher et al. 2005).
%An rapid decline of 0.52 mag in 42 min
%in the B band was observed (Xie et al. 1992).

 \begin{figure*}
   \centering
   \includegraphics[width=10cm]{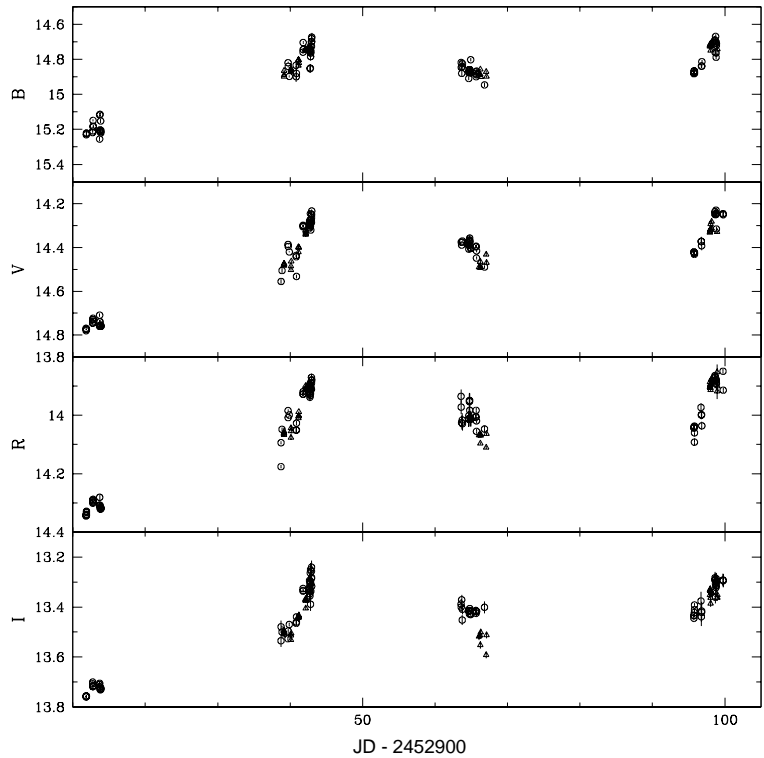}
   \caption{Light curves of 3C 66A in the $B$, $V$, $R$, $I$ bands. The circles
   represent the data from the 1.0 m Mt. Lemmon telescope, and the
   triangles
   are from the 0.6 m Sobaeksan telescope.
   }
              \label{FigGam}%
    \end{figure*}
%However, one paper found the opposite correlation
%(???) in infrared bands (intraday variations).

Our $BVRI$ light curves are shown in Fig. 1. Data obtained from
different observatories are represented by different symbols.
During our observing runs, the overall magnitude variations were
$\Delta B=0.59$, $\Delta V=0.55$, $\Delta R=0.50$, and $\Delta
I=0.52$ over 88 days. Our observations of this object consist of
four sessions: September, October, November, and December 2003.
From Fig. 1, it can be seen that the source stays at a relatively
low state, $R\sim14.3$ in the September session. However, 3C 66A
brightened to $R\sim14.0$ in the following three sessions. It can
also be seen that the source brightens gradually in the October,
and December sessions, but fades in the November session, although
there are fluctuations in each observing run. We find that the
brightest magnitude of October, $R\sim13.8$ is same as that of
December. The time separation of these two epochs is about 57
days. Short-time-scale variations (e.g. intraday variations) were
also searched for during our observing runs, but no significant
such variations were found.

%                                     Two column figure (place early!)
%______________________________________________ Gamma_1 (lg rho, lg e)

%

%\begin{figure*}
%   \centering
%   \includegraphics[width=\textwidth]{vir.eps}
%   \caption{V-I colour versus R mag for our blazar objects.
%   }
%              \label{FigGam}%
%    \end{figure*}

%\subsection{AO 0235+164}

\subsection{PKS 0420$-$01}

This quasar has revealed a strong variability in the optical band
(Webb et al. 1988). A noticeable flare was detected in late 1979,
when a 1.3 mag increase in 5 days was registered, followed by a
1.7 mag decrease in 23 days (Villata et al. 1997). A strong
variability is also evident from the 20 yr light curve (from 1970
to 1990) reported by Smith et al. (1993). Wagner et al. (1995)
noticed flux variations with time scales of the order of 1-10
days. The flare of 1992 February-March was the highest optical
state observed until then ($R=14.6$); in that period EGRET
registered the highest $\gamma$-ray flux density. Since low fluxes
or non-detections at $\gamma$-ray energies correspond to low
optical states, a direct correlation between the optical and
$\gamma$-ray emission was suggested. The most noticeable variation
was a fall of 2.64 mag in 40 days observed from 1995 September 15
to October 25 in the $R$ band (Villata et al. 1997).
Microvariability with $\Delta R=0.12$ mag in 40 minutes was also
detected (Villata et al. 1997).
\begin{figure*}
   \centering
   \includegraphics[width=10cm]{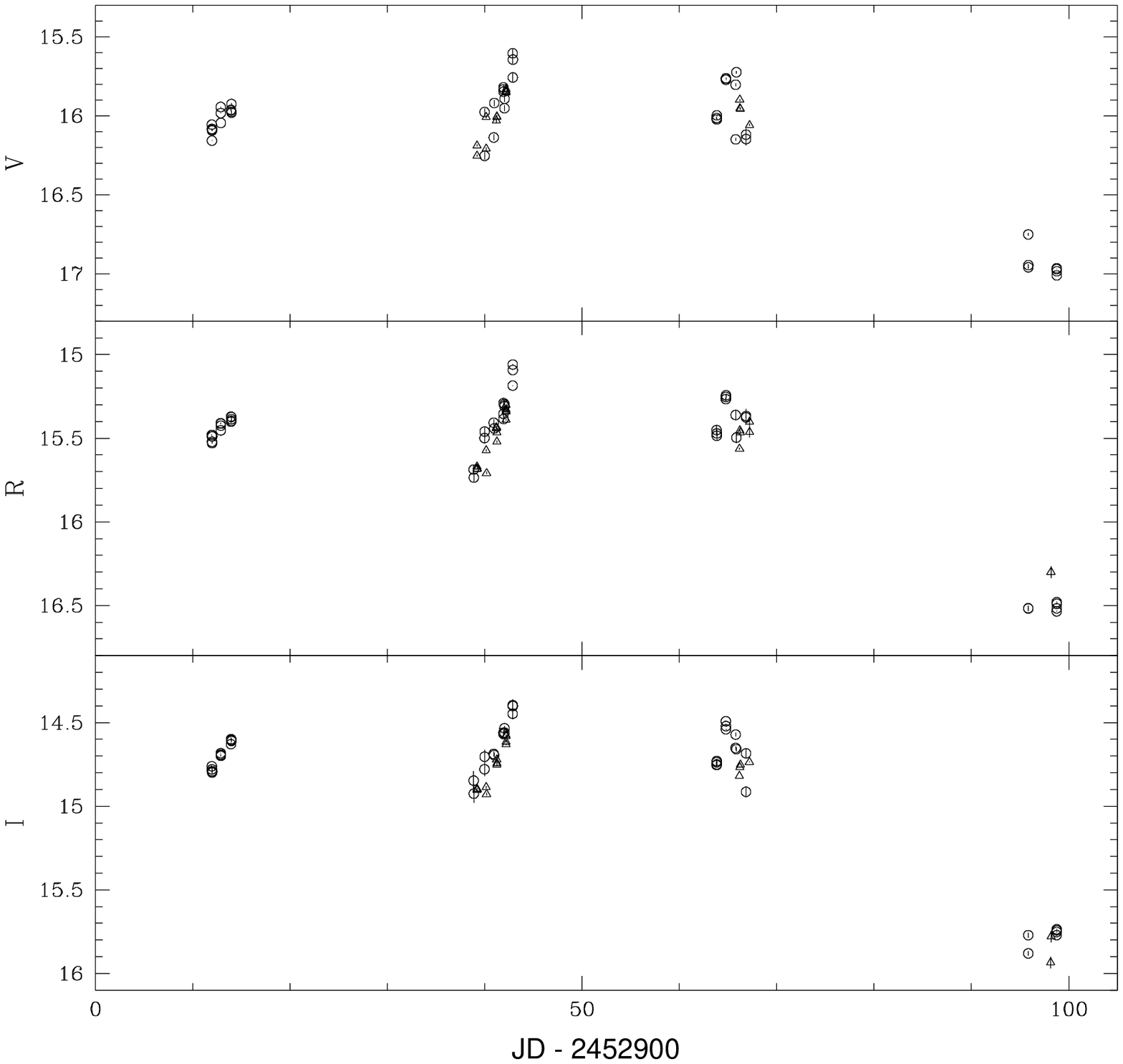}
   \caption{Light curves of PKS 0420$-$01 in the $V,~ R,~ I$ bands. The
   symbols are the same as in Fig. 1.
   }
              \label{FigGam}%
    \end{figure*}

The $VRI$ light curves of PKS 0420$-$01 are shown in Fig. 2.
During our observing runs, the overall magnitude variations were
$\Delta V=1.41$, $\Delta R=1.48$, and $\Delta I=1.54$. The maximum
$R=15.06$ happened on October 30, 2003 (JD=2452942.87), and the
minimum $R=16.54$ was on December 25, 2003 (JD=2452998.76). During
the September run, the light curve over three days (September 29,
30, and October 1) shows that the source was near $R\sim15.4$, and
it slightly brightened over this period, but with only about 0.12
mag variation in this band.
% magnitude brightening from the first night to the last night.
During the October session, the source gradually brightened from
$R=15.74\pm0.03$ on October 26, to $R=15.06\pm0.01$ on October 30,
which is the maximum over the whole monitoring period. The average
brightness during these five days is $R\sim15.4$, which is same as
the average magnitude of the September session. We find that the
source also stays at the same brightness level ($R\sim15.4$) in
the November run, although there are fluctuations from day to day.
The $R$ magnitude was $\sim$15.47 on November 20, and it
brightened to $\sim15.25$ on November 21. On November 22, the
source faded to $R\sim15.43$, and after that, it stayed at
$R\sim15.4$. However, the source had faded by about 1 mag to
$R\sim16.5$ in the December session.

%\subsection{PKS 0528+134}

\subsection{S5 0716+71}

The BL Lac object S5 0716+71 was included in the S5 catalog of the
Strong Source Survey performed at 4.9 GHz (K\"{u}hr et al. 1981).
The largest optical database on S5 0716+71 ever published was
collected and analyzed in Raiteri et al. (2003), which consists of
a total of 4854 $UBVRI$ data points obtained from eight
observatories from 1994 to 2001. Four major optical outbursts were
observed: at the beginning of 1995, in late 1997, at the end of
2000, and in autumn 2001. In particular, an exceptional
brightening of 2.3 mag in 9 days was detected in the $R$ band just
before the BeppoSAX pointing of October 30, 2000. They found that
the long-term trend shown by the optical light curves seems to
vary with a characteristic time scale of about 3.3 years. The
source has been recently monitored at radio and optical
wavelengths by more than 40 telescopes in the northern hemisphere,
during a WEBT campaign lasting from September 2003 to June 2004.
Nesci et al. (2002) presented a study of the optical intraday
variability (IDV) of S5 0716+71 over 52 nights. They found typical
variation rates of 0.02 mag per hour, and a maximum rising rate of
0.16 mag per hour. S5 0716+71 has been observed with $INTEGRAL$ on
2-6 April 2004 (Pian et al. 2005). It was detected with IBIS/ISGRI
up to 60 keV, with a flux of
$\rm\sim3\times10^{-11}~erg~s^{-1}~cm^{-2}$ in the 30-60 keV
interval.

S5 0716+71 was the most extensively monitored source in our
observations. The observations span about 140 days, occurring in
September, October, November, December 2003, and February 2004.
The light curves in the Johnson's $BV$ and Cousins' $RI$ bands are
shown in Fig. 3, where the different symbols refer to the various
telescopes listed in Table 1. It can be seen from Fig. 3 that the
source exhibited strong variability, with similar trends (but with
slightly different amplitudes) in all bands. The overall magnitude
variations were $\Delta B=1.42$, $\Delta V=1.37$, $\Delta R=1.37$,
$\Delta I=1.36$, where the differences are mainly due to the
different temporal samplings. During our observing runs, the
maximum $R$-band brightness, $R=12.62$, was recorded on 10
December 2003 (JD=2452983.8756). We note that a rapid brightness
increase was detected during the 4 days of 19 -23 November 2003,
and which is shown in Fig. 4. The overall variation in the $R$
band was $\Delta R=1.15$, from $R=13.94\pm0.01$ on JD=2452962.916
to $R=12.79\pm0.01$ on JD=2452966.854, both observed from the Mt.
Lemmon Observatory.

\begin{figure*}
   \centering
   \includegraphics[width=10cm]{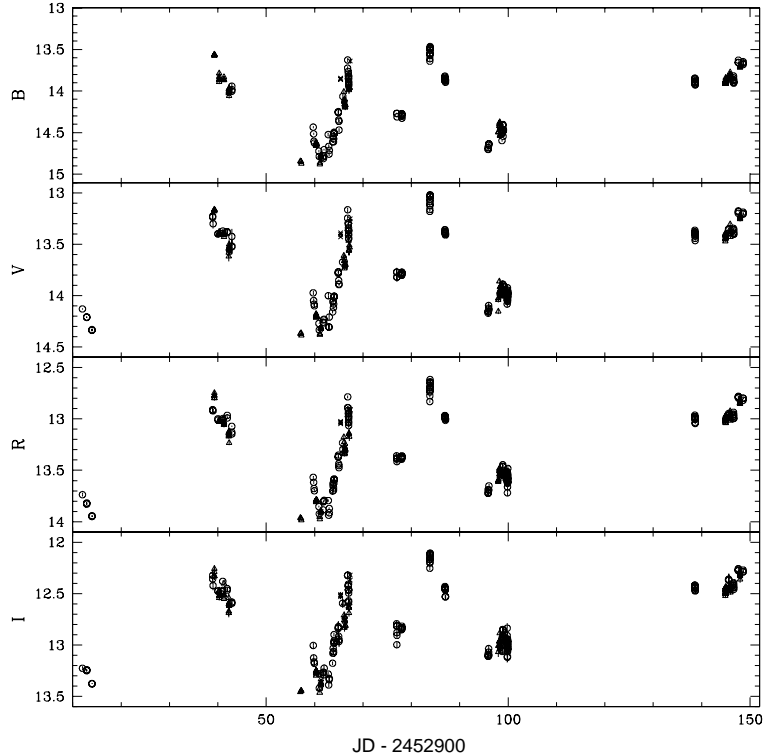}
   \caption{Light curves of S5 0716+71 in the $B,~ V,~ R,~ I$ bands. The
   circles and triangles are the same as in Fig. 1, and the crosses
   are the data from the 1.8 m Bohyunsan telescope.
   }
              \label{FigGam}%
    \end{figure*}

\begin{figure*}
   \centering
   \includegraphics[width=10cm]{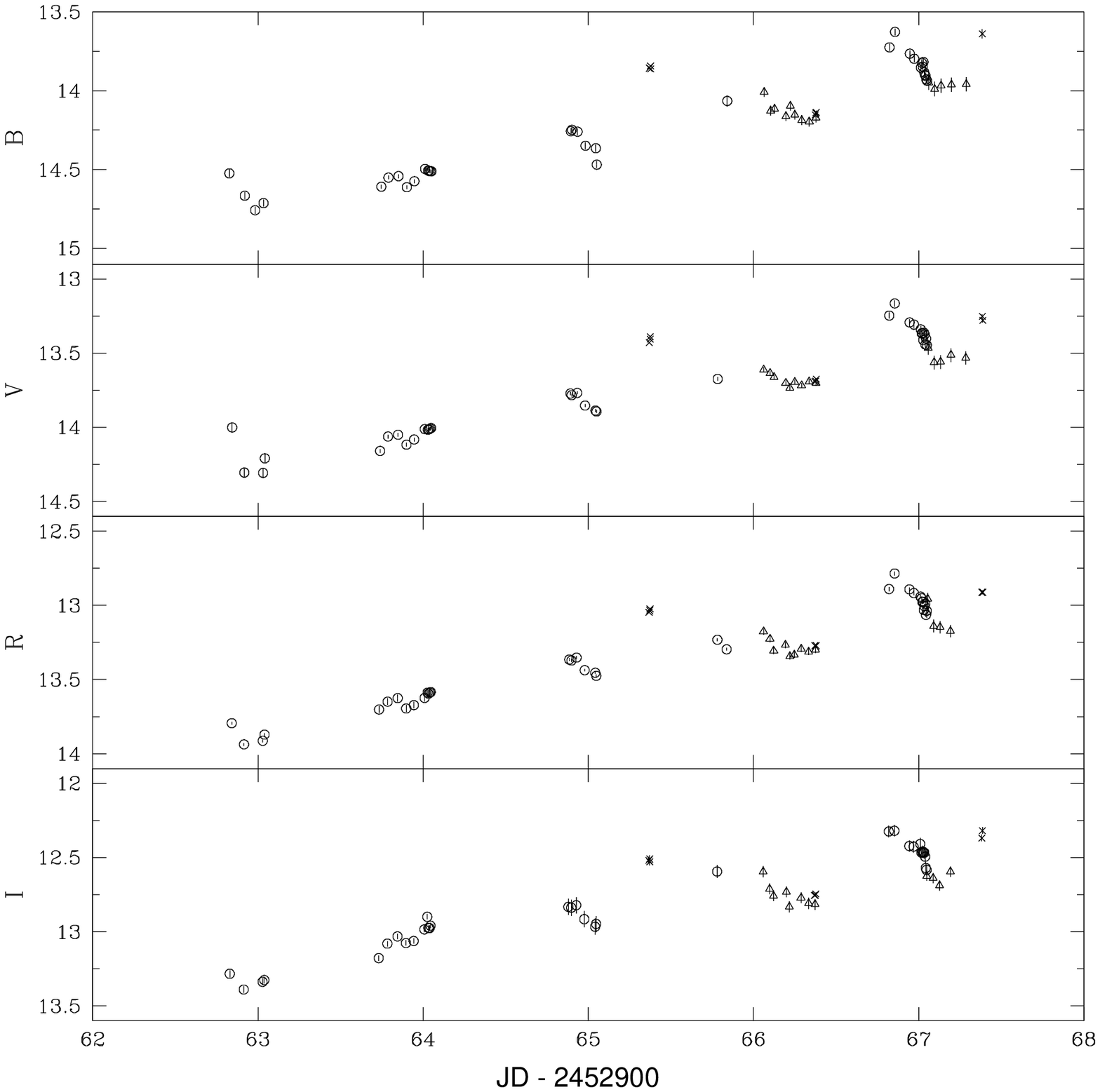}
   \caption{The variability of S5 0716+71 during 19 - 23 November
   2003, in the $B$, $V$, $R$, $I$ bands. The symbols are the same as
   in Fig. 3.
   }
              \label{FigGam}%
    \end{figure*}

Interestingly, we find that the shape of the profile of this rapid
brightness increase is significantly linear. We have fitted
least-square regression lines to the light curves in the $BVRI$
bands obtained from all three telescopes.
%These best-fit lines are shown in Fig. 4 and
The slopes of the best-fit lines are given in Col. 2 of Table 3,
together with the regression coefficients for the best fits in
Col. 3.
%from Lemmon Observatory only (column 2 and 3) and data from
%all observatories (column 4 and 5), respectively.
It is noted that the slopes are essentially the same for the
different passbands. The differences in the slopes are always
$\leq 1\sigma$, and are therefore not statistically significant.
Our results show that the rapid brightening can be described by
linear trends on a magnitude scale (which, therefore, corresponds
to exponential intensity variations), and this rapid brightening
rate was $\sim 0.2$ mag per day over 4 days. However, it should be
noted that this rapid brightening is just a mean behaviour, and
may also be due to the lack of data in the gaps.

\begin{table*}
\caption{Slopes and regression coefficients ($r$) of the
least-square linear fits to the $BVRI$ magnitude variations of S5
0716+71, during 19-23 November 2003.}
\begin{tabular}{lcc}
\hline \hline
Filters & slope$\pm\sigma$ & $r$  \\
\hline

$B$ &   -0.197$\pm$0.013 & -0.90 \\
$V$ &   -0.194$\pm$0.013 & -0.90 \\
$R$ &   -0.198$\pm$0.012 & -0.92 \\
$I$ &   -0.179$\pm$0.012 & -0.89 \\

\hline
\end{tabular}

\end{table*}

%It is noted that most of the data are obtained from the Mt. Lemmon
%Observatory. We thus fitted least-square regression lines to the
%data only from that observatory to avoid the possible systematic
%difference in different telescopes. The slopes in each band are
%shown in column (4) of Table 2, and the regression coefficients
%are in column (5). It is apparent that the regression coefficients
%for the linear fits of the data from only Mt. Lemmon observatory
%are $\sim-0.98$ in all bands, which is stronger than that of all
%observatories.

%But our sampling is not dense. It is worthwhile to investigate how
%the data sampling would influence our results.
%Actually, one can
%see that the regression coefficients for the linear fits of the
%data from only Lemmon Observatory are $>95\%$ in all passbands,
%whereas the regression coefficients from all observatories
%(temporal sampling is denser) are smaller, $\sim90\%$, but still
%significant. The slopes are consistent between two cases within
%the error. So at least for our observations, the data sampling may
%not influence the results within the error.

Rapid variations of the flux of S5 0716+71 have been observed on
different occasions (e.g. Ghisellini et al. 1997; Sagar et al.
1999; Xie et al. 2004). In our observations, we also searched for
intraday variations. Variations, with amplitudes of $>$0.15 mag
over a few hours in one night, were found on November 23 and
December 10, 2003, and which are shown in Fig. 5, and Fig. 6,
respectively. Both of them were detected with the Mt. Lemmon
Observatory. The lower panels in these figures show the difference
between the instrumental magnitude of the two field standard
stars. In Fig. 5, it can be seen that the source brightened from
$V=13.25\pm0.02$ ($R=12.89\pm0.01$) to $V=13.17\pm0.02$
($R=12.79\pm0.01$) within 48 min, and afterwards it faded to
$V=13.45\pm0.02$ ($R=13.06\pm0.01$) in the remaining observing
time ($\sim4.6$ hr) for this night. The overall magnitude
variations were $\Delta V=0.28$ mag ($\Delta R=0.27$ mag) during
this night. The variability parameters C (introduced by Romero,
Cellone \& Combi 1999; if C$>$2.576, then the source is variable)
are 3.64 ($V$ band), and 5.27 ($R$ band). The lower panels show
that the maximum deviation of the two field standard stars is
$\Delta V_{\rm max}=0.07$ mag in the $V$ band, and $\Delta R_{\rm
max}=0.06$ in the $R$ band.
%Since the overall intraday variations
%have amplitudes larger than 3 $\Delta V_{\rm max}$, and 3 $\Delta
%R_{\rm max}$, respectively, we can be reasonably confident that
%they are real.
%That is, the
%confidence of these intraday variations are larger than 3$\sigma$.
In Fig. 6, the source brightened from $V=13.18\pm0.01$
($B=13.64\pm0.01$) to $V=13.02\pm0.01$ ($B=13.47\pm0.01$) within
about 1.4 hr, then faded. The rising rate, of 0.002 mag per minute
in both $B$ and $V$ bands, was found over the duration of the
first 1.4 hr. This is in good agreement with the results of
Villata et al. (2000), where there was a monotonic brightness
increase from $B=14.13\pm0.01$ to $B=13.98\pm0.01$ in about 130
minutes, with the steepest (linear) part having a rising rate of
0.002 mag per minute, and a duration of about 45 minutes.
Moreover, the authors suggested that such a gradient seems to be
typical of fast variations, with no evident difference between
rise and fall.

%                                     Two column figure (place early!)
%______________________________________________ Gamma_1 (lg rho, lg e)
%

\begin{figure*}
   \centering
   \includegraphics[width=10cm]{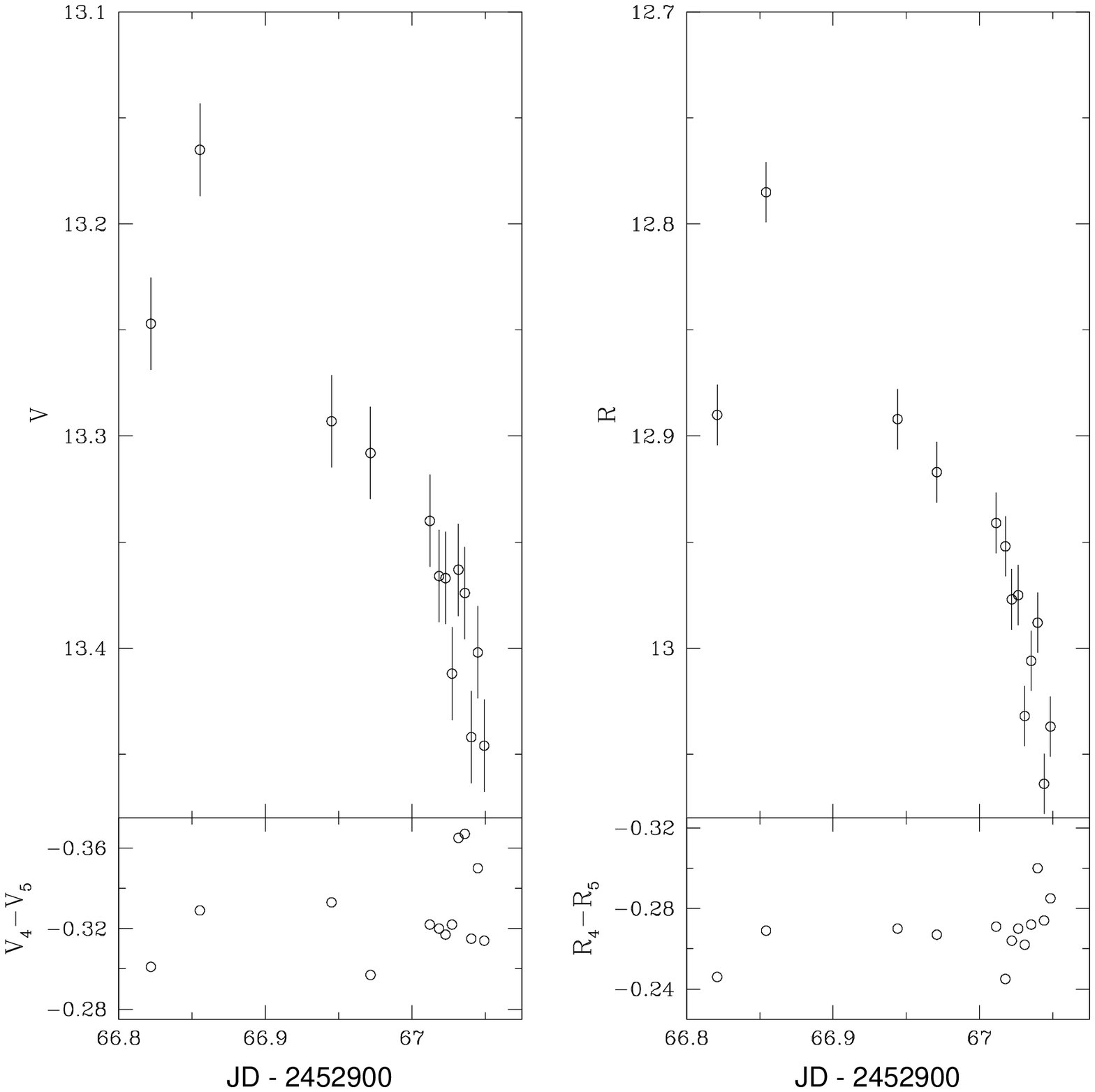}
   \caption{Intraday variations of S5 0716+71 on 23 November
   2003, in the $V$ band (left panel) and in the $R$ band (right panel).
    All data are from the 1.0 m Mt. Lemmon telescope.
   }
              \label{FigGam}%
    \end{figure*}

\begin{figure*}
   \centering
   \includegraphics[width=10cm]{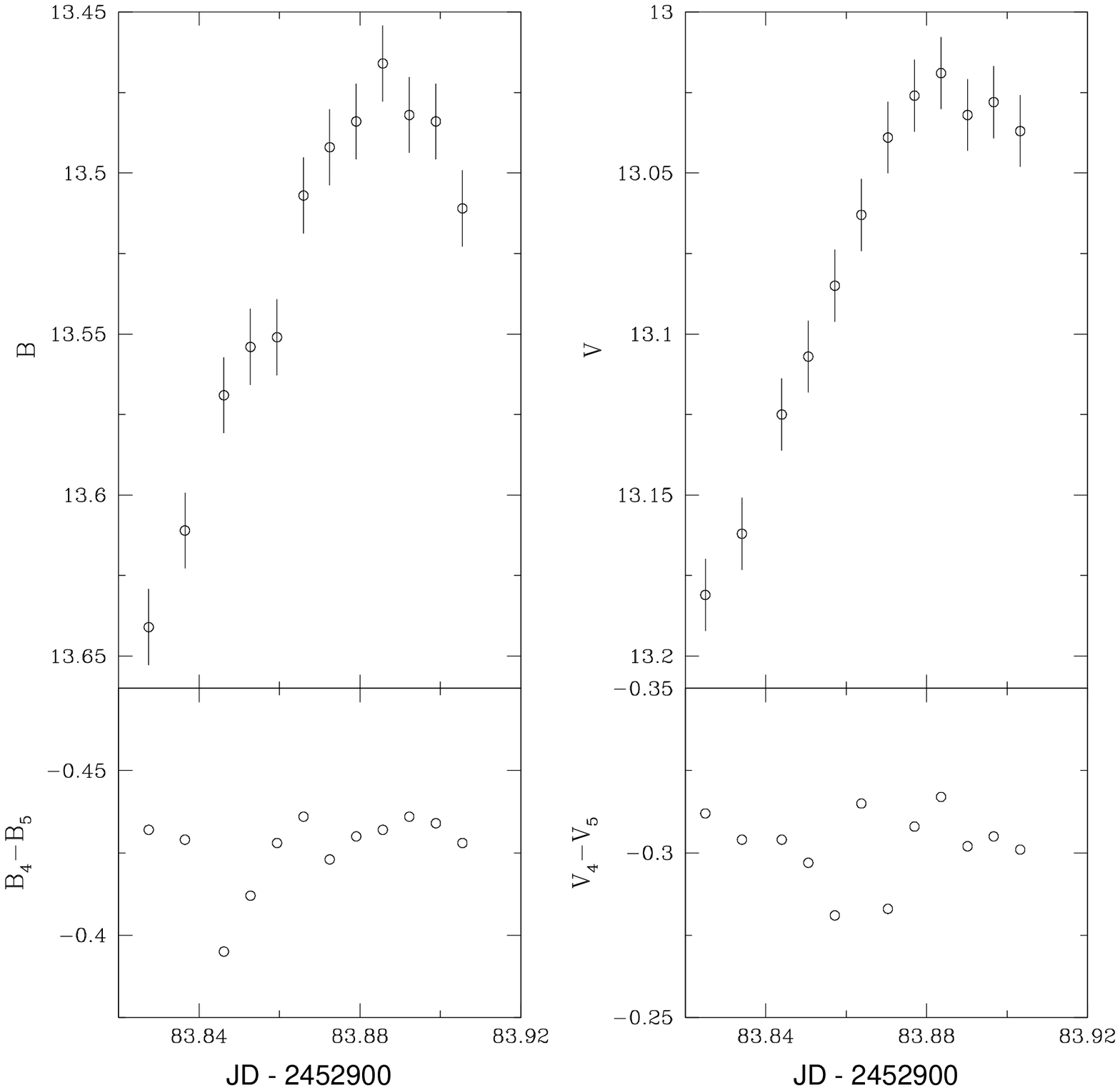}
   \caption{Intraday variations of S5 0716+71 on 10 December
   2003, in the $B$ band (left panel) and in the $V$ band (right panel).
    All data are from the 1.0 m Mt. Lemmon telescope.
   }
              \label{FigGam}%
    \end{figure*}

\subsection{PKS 0735+17}

PKS 0735+17 was first classified as a BL Lac object by Carswell et
al. (1974). It is both radio (K\"{u}hr et al. 1981) and X-ray
selected (Elvis et al. 1992). Falomo \& Ulrich (2000) gave a lower
limit on the object's redshift, of $z>0.5$, and Xie et al. (2002)
have estimated its black hole mass. Tommasi et al. (2001) observed
the optical polarimetry of PKS 0735+17 in 1999 December, and
gathered together previous polarimetric data. The source has shown
very different levels of polarization percentage in past years,
from around 1\% to more than 30\%. PKS 0735+17 is also known as an
optical and infrared intraday variable blazar (Massaro et al.
1995; Heidt \& Wagner 1996; Bai et al. 1998). With observational
data spanning over 90 yr, the long-term periodicity of this blazar
has been investigated; its optical variability shows the
possibility of a 4.89 yr period (Webb et al. 1988; Smith et al.
1988), and also a 14.2 yr period (Fan et al. 1997).

\begin{figure*}
   \centering
   \includegraphics[width=10cm]{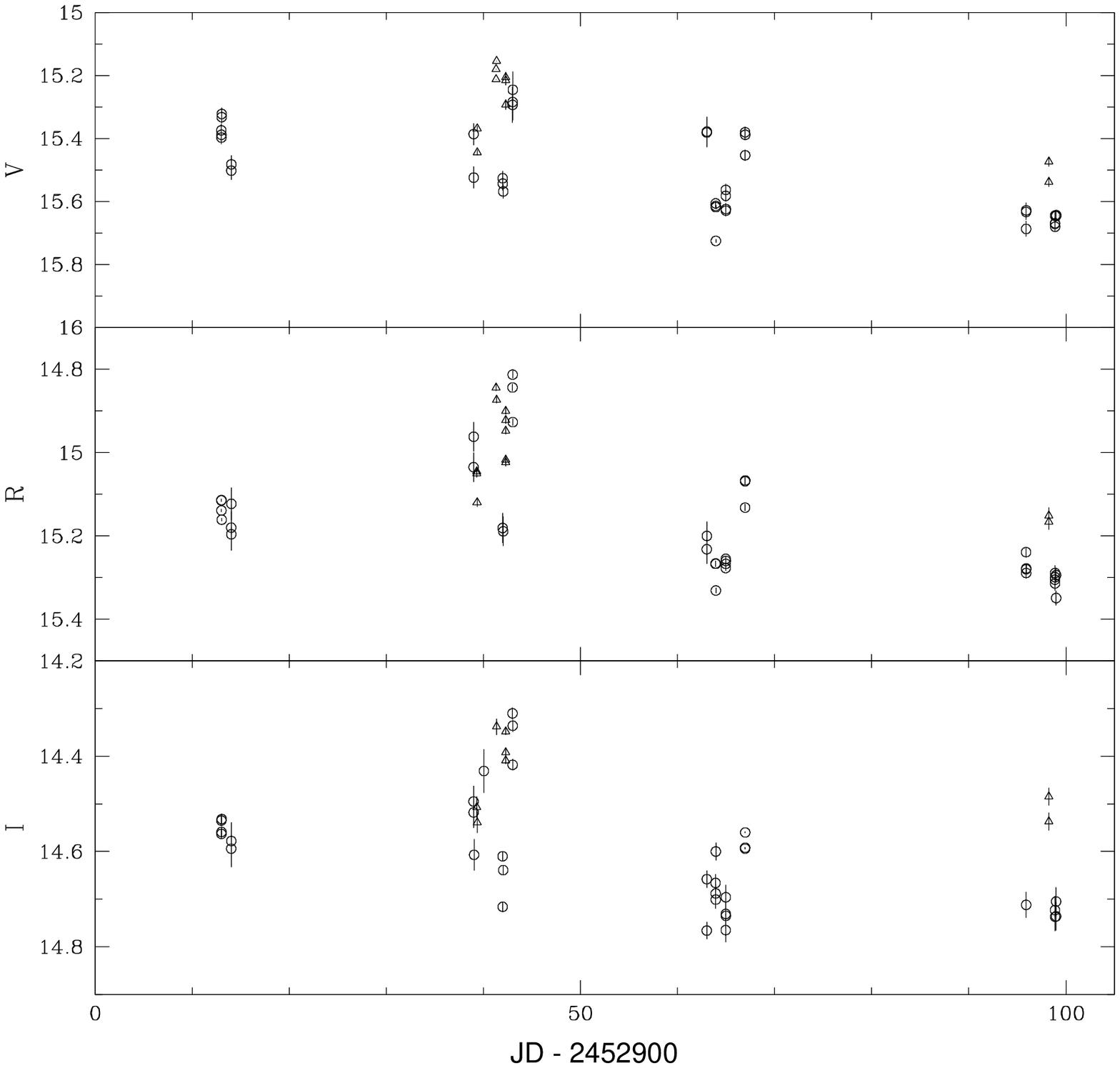}
   \caption{Light curves of PKS 0735+17 in the $V,~ R,~ I$ bands. The
   symbols are the same as in Fig. 1.
   }
              \label{FigGam}%
    \end{figure*}

Our $VRI$ light curves are shown in Fig. 7. During our observing
runs, the overall magnitude variations were $\Delta V=0.57$,
$\Delta R=0.54$, and $\Delta I=0.46$. From Fig. 7, it can be seen
that this source is variable almost all the time, although the
overall variations are not relatively large. This source was
relatively stable from September 30 to October 1, 2003. The
average $R$ magnitude of PKS 0735+17 was $15.13\pm0.02$ on
September 30, and $15.17\pm0.04$ on October 1. From October 26 to
October 30, the apparent fluctuations are present in all bands
(see Fig. 7). On October 30, the maximum $R$ brightness of whole
monitoring period was reached at $14.81\pm0.01$ (JD=2452942.998).
In the $R$ band, a brightening from $15.26\pm0.01$ on November 21
to $15.09\pm0.04$ on November 23 was detected. One month later,
the object was observed at $R=15.27\pm0.02$ on December 22, and
brightened to $R=15.16\pm0.02$ after two days. Finally, it faded
to $R=15.31\pm0.02$ on December 25.

\subsection{OJ 287}

OJ 287 is one of the most extensively observed BL Lac objects.
%The
%redshift is 0.306 (Sitko \& Junkkarinen 1985), and the host galaxy
%has been marginally resolved (Heidt et al. 1999).
It has been observed for over 100 years, providing a very good
historical light curve. In 1972, it reached maximum light, with
$V\sim12$. Pursimo et al. (2000) presented intensive optical,
infrared, and radio monitoring of OJ 287, taken between 1993 and
1998. These monitoring results show that the optical and infrared
light curves displayed continuous variability on timescales
ranging from tens of minutes to years.
%Optical polarization also showed large random
%variability. In the radio bands, superluminal motion was first
%observed from VLBI polarization data at $\lambda=6$cm (Roberts,
%Gabuzda, \& Wardle 1987). Radio light curves can be found in many
%papers (e.g., Tateyama et al. 1999; Venturi et al. 2001).
%The periodicity of OJ 287 is of great interest.
The historical light curve shows several outbursts with a
recurrence period of about 11.65 yr (Sillanp\"{a}\"{a} et al.
1988a). The 12 yr cycle was discussed in many occasions (e.g.
Villata et al. 1998b; Abraham 2000; Katajainen et al. 2000;
Valtaoja et al. 2000).
%This possible periodicity was
%also studied by Kidger, Takalo, \& Sillanp\"{a}\"{a} (1992),
%Kidger (2000), and Fan et al. (2002), who obtained similar
%results.
Sillanp\"{a}\"{a} et al. (1988a) modeled these periodic outbursts
by using a binary supermassive black hole system, and predicted
that the next outburst would occur during late 1994. This
prediction was confirmed by several studies (Kidger et al. 1995;
Sillanp\"{a}\"{a} et al. 1996a, 1996b; Arimoto et al. 1997; Fan et
al. 1998; Jia et al. 1998; Pursimo et al. 2000; Efimov et al.
2002).
\begin{figure*}
   \centering
   \includegraphics[width=10cm]{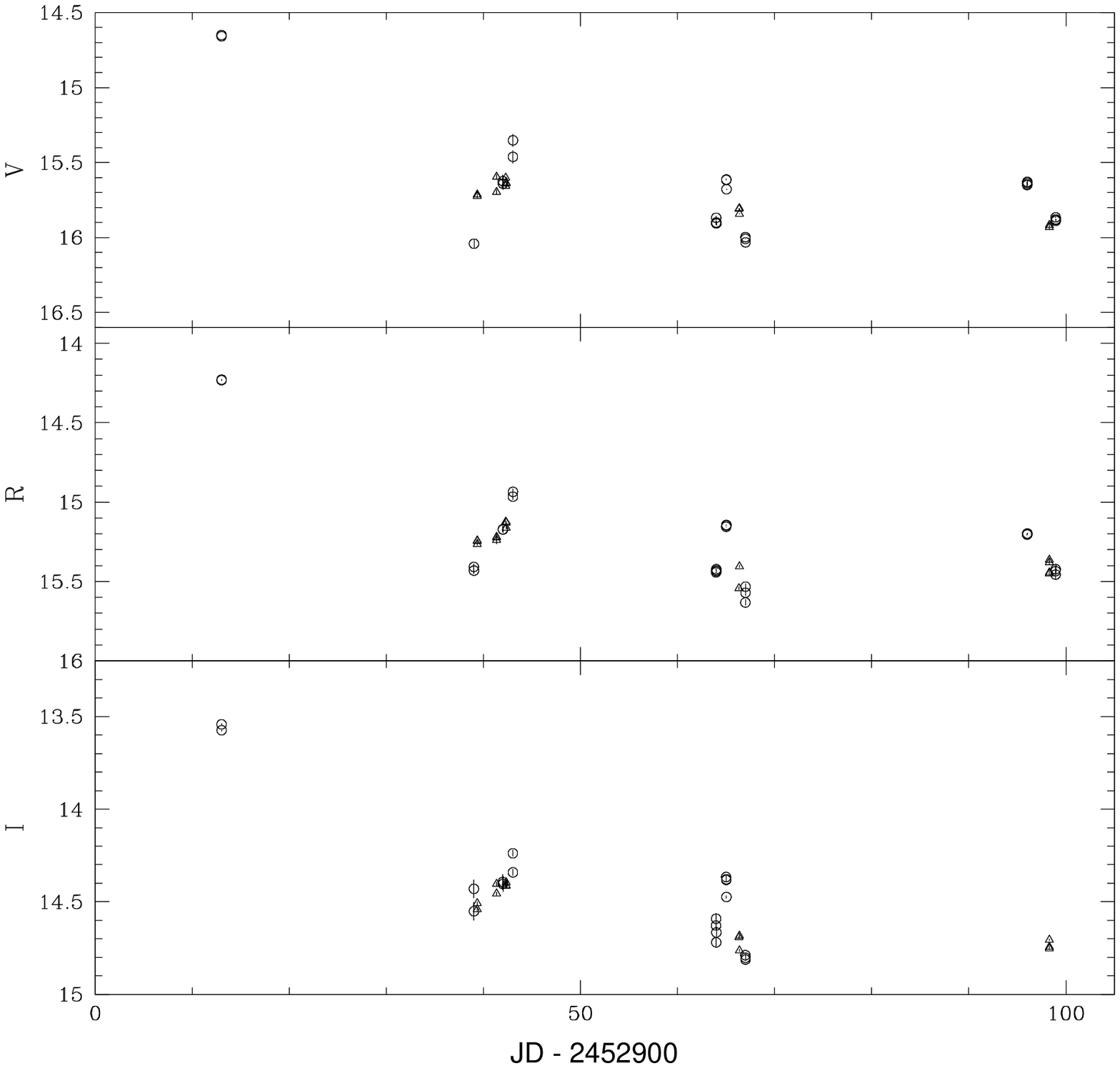}
   \caption{Light curves of OJ 287 in the $V,~ R,~ I$ bands. The
   symbols are the same as in Fig. 1.
   }
              \label{FigGam}%
    \end{figure*}

Our $VRI$ light curves are shown in Fig. 8. During our observing
runs, the overall magnitude variations were $\Delta V=1.39$,
$\Delta R=1.40$, and $\Delta I=1.27$. The maximum $R=14.23\pm0.01$
occurred on September 30 (JD=2452913.001), whereas the minimum
$R=15.63\pm0.02$ was on November 23 (JD=2452966.982). The source
was observed at $R=14.23\pm0.01$ ($V=14.65\pm0.03$,
$I=13.55\pm0.05$) on September 30, 2003, which was recorded at the
Mt. Lemmon Observatory. About one month later, it faded to
$R=15.42\pm0.02$ on October 26. After that, it gradually
brightened, and reached $R=14.95\pm0.02$ on October 30. In
November, the source brightness stayed at a similar level as that
in October. On November 20, the $R$ magnitude was recorded at
$15.43\pm0.01$, but it brightened to $15.15\pm0.01$ on November
21. Later, the source faded to $15.58\pm0.02$ on November 23. In
December, a fading from $R=15.20\pm0.01$ on December 22, to
$R=15.44\pm0.03$ on December 25 was detected over three days, at
the Mt. Lemmon Observatory.

\subsection{3C 345}

3C 345 was classified as an OVV quasar by Penston \& Cannon
(1970). The typical behaviour for this object has been to show
about 2 mag outbursts, which occur quite frequently. Such large
outbursts happened in 1967/1968, 1971/1972, 1982/1983, and
1991/1992 (Schramm et al. 1993). Sillanp\"{a}\"{a} et al. (1988b)
measured $V=14.56$ in 1982, while Schramm et al. (1993) observed
the $V$-band brightness near 18 mag in 1989. In 1991/1992, a large
2.5 mag outburst was observed (Schramm et al. 1993). Kidger \&
Takalo (1990) observed the historical minimum at $B=18.66$.

\begin{figure*}
   \centering
   \includegraphics[width=10cm]{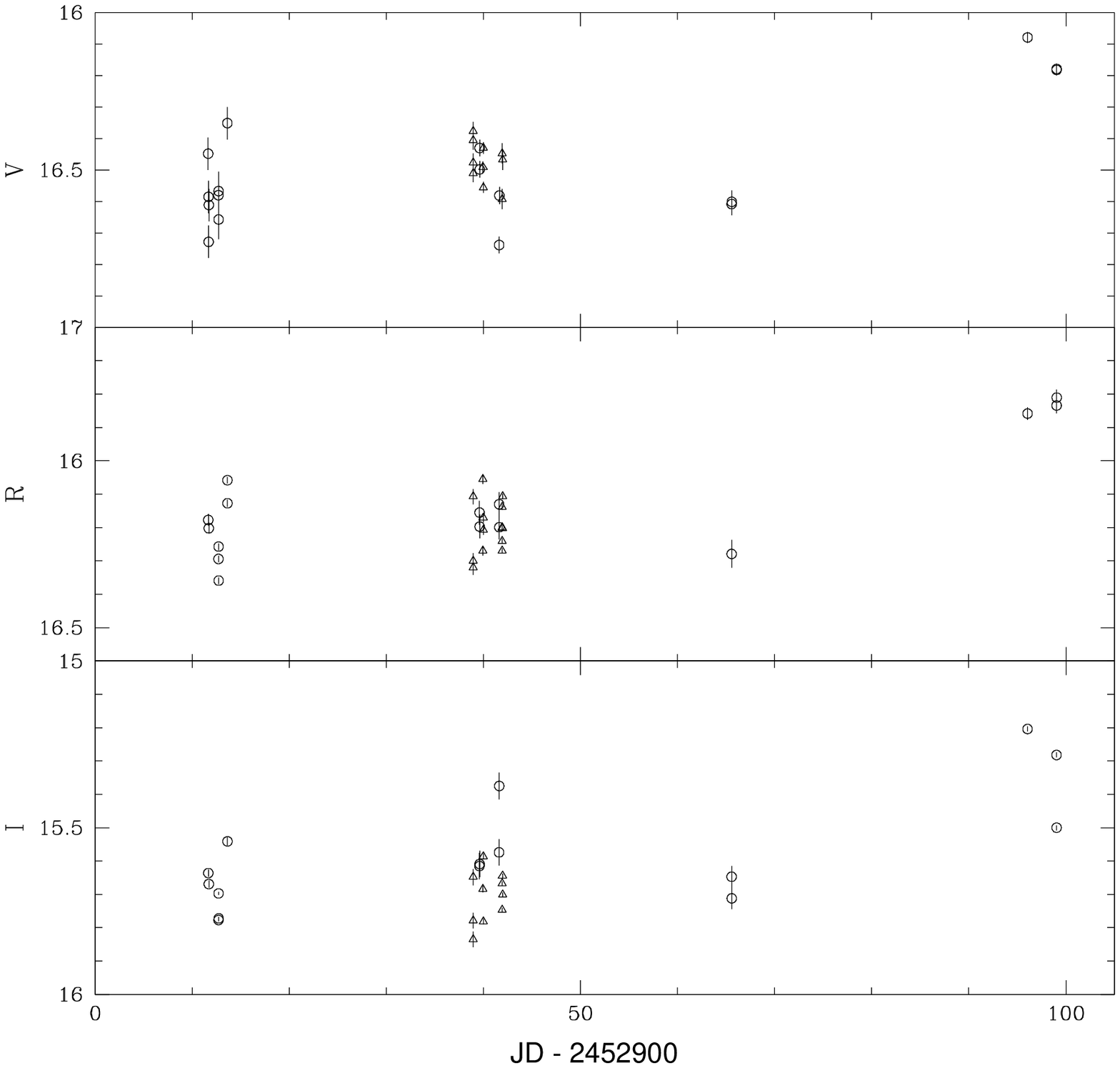}
   \caption{Light curves of 3C 345 in the $V,~ R,~ I$ bands. The
   symbols are the same as in Fig. 1.
   }
              \label{FigGam}%
    \end{figure*}

The $VRI$ light curves are shown in Fig. 9. During our observing
runs, the overall magnitude variations were $\Delta V=0.66$,
$\Delta R=0.55$, and $\Delta I=0.63$.
%For this source, we have
%only a few observations.
The maximum $R$ brightness $R=15.81\pm0.02$ was observed on
December 25 (JD=2452999.055), and the minimum $R=16.36\pm0.01$ was
on September 30 (JD=2452912.710). Variations with small amplitudes
on time scales of days are found in all bands, in September,
October, and December 2003.

\subsection{BL Lac}

BL Lac is the prototype of the BL Lac object class of AGNs, and
one of the best studied blazars. Shen \& Usher (1970) have
investigated its historical light curves, and found a range of
variation of 4.2 magnitude in the $V$ band, and a strong outburst
in which BL Lac varied over nearly its entire range (12.4$\leq B
\leq$16.7 mag) during 400 days. Investigations of its long-term
variability have been done by Webb et al. (1988), and Carini et
al. (1992), which show that the faintest magnitudes were $B=17.99$
mag, and $V=16.73$ mag.
%Recent WEBT papers (Villata et al.
%2002, 2004).
Recently, Villata et al. (2004) presented $UBVRI$ light curves
obtained by the WEBT from 1994 to 2002, including the last,
extended BL Lac 2001 campaign. Their analysis of the colour
indices reveals that the flux variability can be interpreted in
terms of two components: longer-term variations occurring on a
few-day time scales appear as mildly-chromatic events, while a
strong bluer-when-brighter chromatism characterizes very fast
(intraday) flares. They suggested that Doppler factor variations
on a ``convex" spectrum could be the mechanism accounting for both
the long-term variations, and their slight chromatism.

\begin{figure*}
   \centering
   \includegraphics[width=10cm]{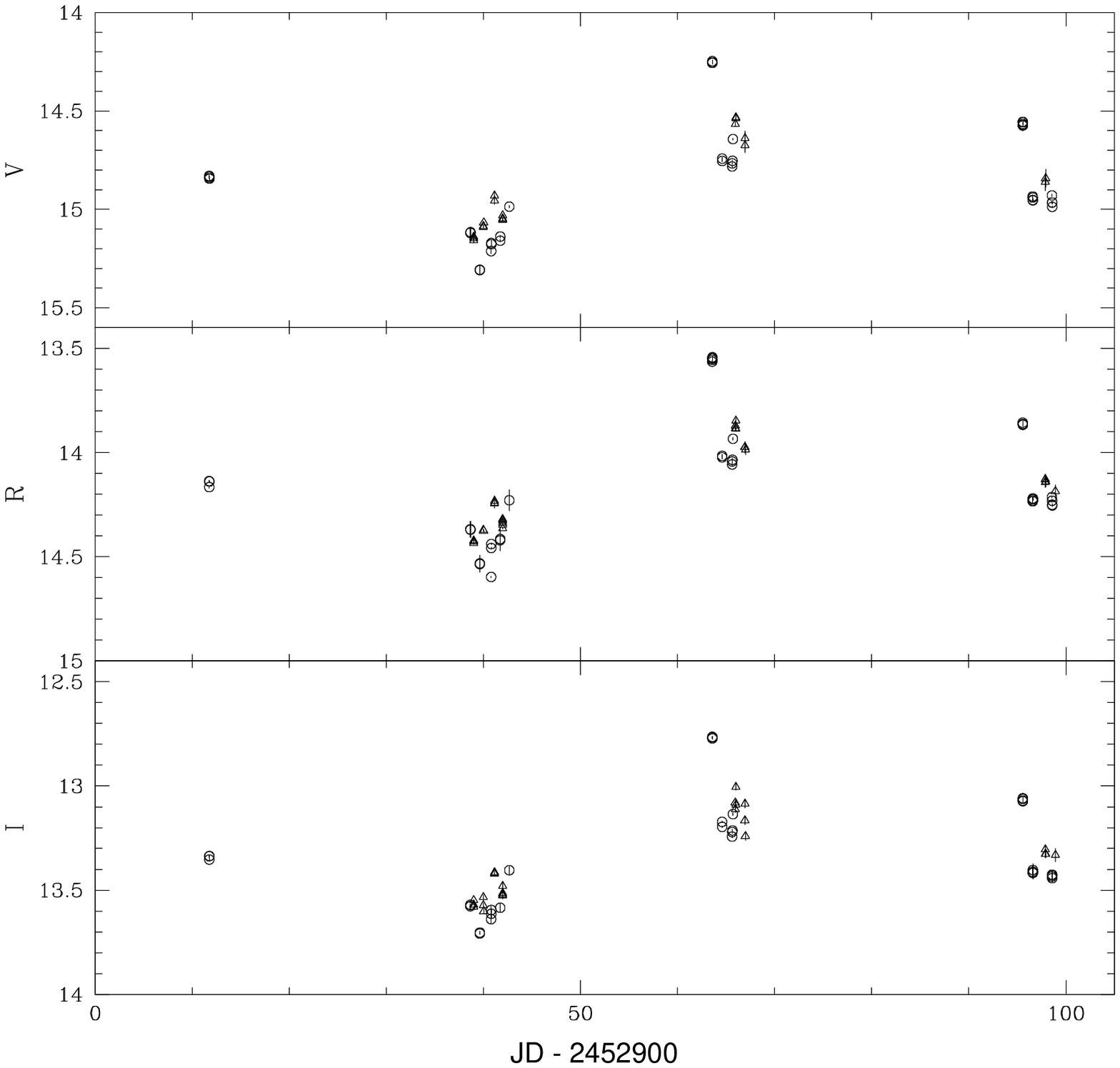}
   \caption{Light curves of BL Lac in the $V,~ R,~ I$ bands. The
   symbols are the same as in Fig. 1.
   }
              \label{FigGam}%
    \end{figure*}

Our $VRI$ light curves are shown in Fig. 10. During our observing
runs, the overall magnitude variations were $\Delta V=1.06$,
$\Delta R=1.05$, and $\Delta I=0.94$. The maximum $R$-band
brightness was recorded at $R=13.55\pm0.01$ on November 20
(JD=2452963.580), whereas the minimum was detected at
$R=14.60\pm0.01$ on October 28 (JD=2452940.784). BL Lac was
observed at $R\sim14.15\pm0.01$ on September 29. It was slightly
fainter in late October. On October 26, the $R$ magnitude was
$R=14.37\pm0.04$. It became $R=14.23\pm0.05$ after four days,
during which small amplitude fluctuations ($\leq0.2$ mag) were
detected from day to day. In the November run, BL Lac first
brightened to $R=13.55\pm0.01$ on November 20
(JD$\sim$2452963.580), then it faded to $R=14.02\pm0.01$ in one
day. On November 22, a brightening of $\sim0.10$ mag in the $R$
band (0.11 mag in the $V$ band, and 0.08 mag in the $I$ band),
from $R=14.03\pm0.01$ at JD=2452965.648 to $R=13.93\pm0.01$ at
JD=2452965.696, was detected in about 69 minutes. The $R$
magnitude was recorded at $13.86\pm0.01$ on December 22, but it
dropped to $14.23\pm0.01$ after one day. After that, it brightened
slightly to $14.14\pm0.02$ on December 24. Finally, it reached
$14.24\pm0.02$ on December 25.

\subsection{3C 454.3}

3C 454.3 is a GeV $\gamma$-ray source, and one of the few blazars
that have been detected by COMPTEL (Blom et al. 1995). The light
curve of 3C 454.3 in the $B$ band from 1966 to 1979 is given by
Lloyd (1984). Data in the $B$ band from 1971 to 1985 are reported
by Webb et al. (1988), and a flare with a variation of 1.28 mag in
63 days was reported in autumn 1979. From 1986 November to 1987
January, the average $B$ magnitude of the source was 16.56, with a
variation of 0.92 mag over 44 days (Corso et al. 1988). The
results of the monitoring program of Villata et al. (1997) show
that a fast variation of 0.27 mag in 2.6 hr occurred in the $R$
band; and that in any band this source has presented only small,
short-term variations, within a total range of less than 0.4 mag
(Villata et al. 1997). The $BVR$ light curves of 3C 454.3 from
November 1994 to December 1997, were presented by Villata et al.
(2001).

\begin{figure*}
   \centering
   \includegraphics[width=10cm]{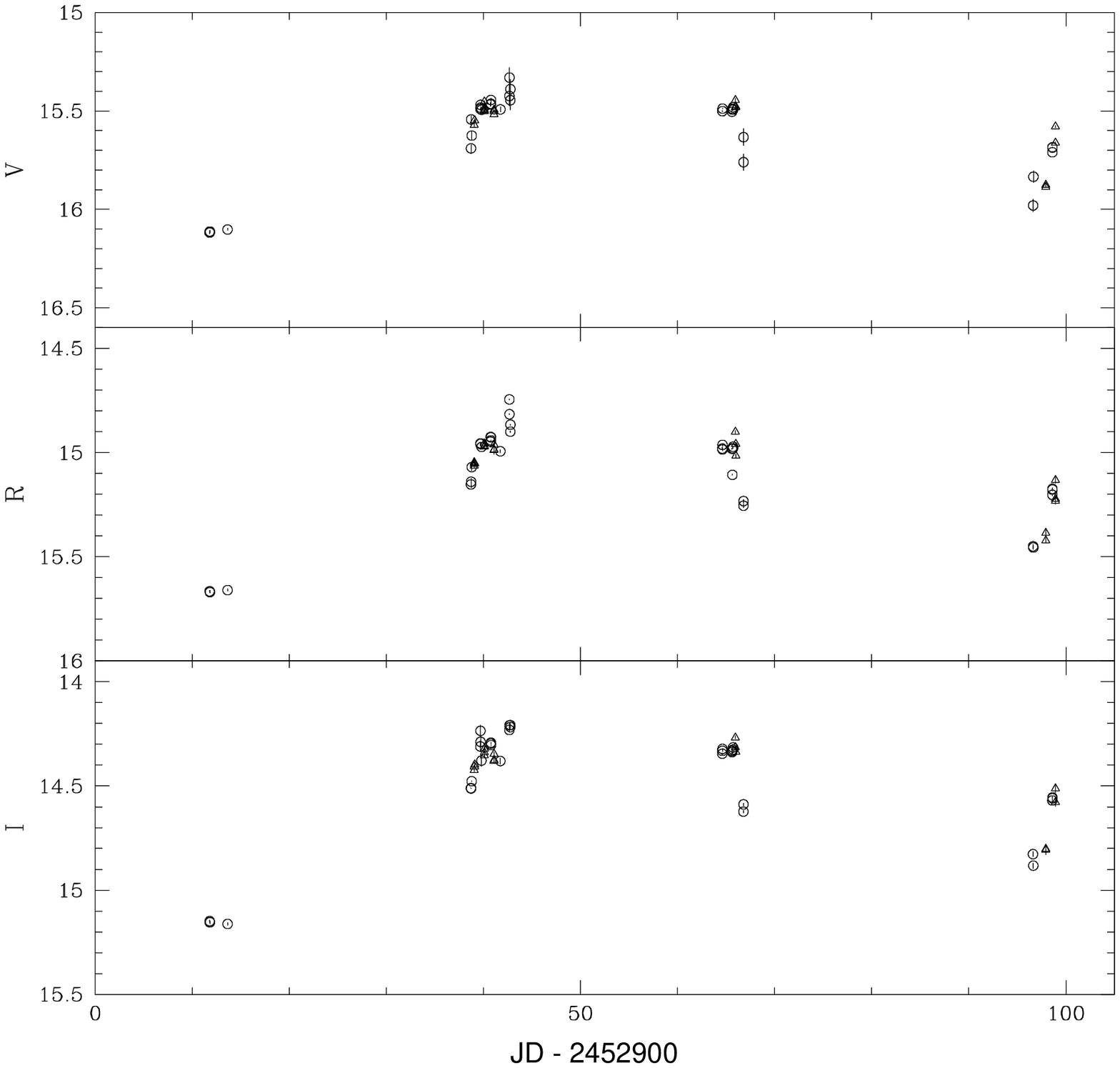}
   \caption{Light curves of 3C 454.3 in the $V,~ R,~ I$ bands. The
   symbols are the same as in Fig. 1.
   }
              \label{FigGam}%
    \end{figure*}

The $VRI$ light curves are shown in Fig. 11. During our observing
runs, the overall magnitude variations were $\Delta V=0.79$,
$\Delta R=0.93$, and $\Delta I=0.95$. The maximum $R$ magnitude
$14.75\pm0.01$ was found on October 30 (JD=2452942.662), while the
minimum $15.67\pm0.01$ was on September 29 (JD=2452911.810). On
September 29 and October 1, the source was stable at
$R=15.67\pm0.01$ ($V=16.11\pm0.01$, $I=15.15\pm0.01$). The object
became brighter in late October. It was detected at $R\sim15.1$ on
October 26, and it gradually brightened over the following four
days, until it reached $R\sim14.8$ on October 30. On November 21
and 22, it was stable at $R\sim15.0$, but faded to $R\sim15.2$ on
November 23. In December, it started at $R=15.45\pm0.01$ on
December 23, then brightened slightly on December 24. Finally, an
$R$ magnitude of $15.13\pm0.02$ was reached on December 25, as
recorded by the Sobaeksan Observatory.

\section{Spectral variability}

In this section, we investigate the relationship between the
spectral changes and the flux variations. The plots of the colour
indices $V-I$ and $R$ magnitudes for each object are given in Fig.
12. In most cases, the colour indices were calculated by coupling
the data taken by the same instrument within 10 min intervals. In
Table 4, we report the principal results of our research. Column 1
gives the object name, Col. 2 the variability range in $R$
magnitude, Col. 3 the number of data points used to calculate the
colour indices $V-I$, Col. 4 the linear Pearson correlation
coefficient between the $V-I$ and $R$ magnitudes, and Col. 5 the
probability that no correlation is present.
% for rejecting the null hypothesis of no correlation.

\begin{figure*}
   \centering
   \includegraphics[width=10cm]{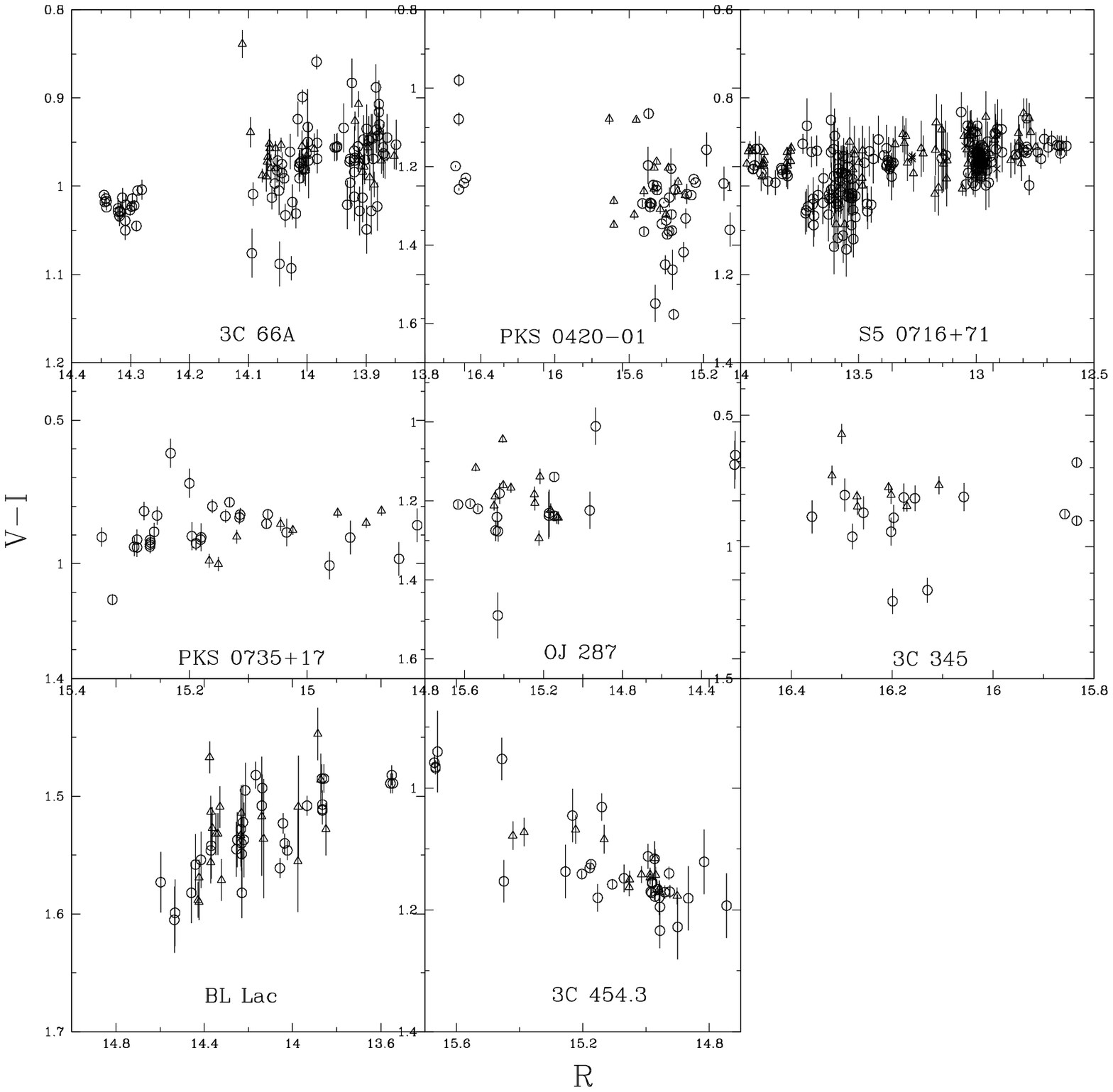}
   \caption{The $V-I$ colour indices versus $R$ magnitudes for the eight
   blazars. The rms colour magnitude errors are indicated by the
   vertical lines. The symbols are the same as in Fig. 3.
   }
              \label{FigGam}%
    \end{figure*}

From Fig. 12, it can be seen that there are no uniform trends of
the colour index changes with the source brightnesses, and the
significance of the dependence of the variation of the colour
index on the source brightness (e.g. as shown by the correlation
coefficient $r$) is not related with the overall variability range
in $R$ magnitude, nor with the number of data points.
Interestingly, we find that the colour indices of two out of three
FSRQs, PKS 0420$-$01, and 3C 454.3, tend to be redder when the
sources are brighter.
%Same as other sources, the V-I color index changes are
%investigated against time and R brightness in Fig. 19, where the
%temporal evolution of R magnitude is also presented.
Especially, we find a significant anti-correlation between the
$V-I$ colour index and $R$ magnitude for 3C 454.3. The linear
Pearson correlation coefficient is found to be $r=-0.837$, with a
probability of $1.07\times10^{-16}$ that no correlation is
present.
% for rejecting the null hypothesis of no correlation.
The anti-correlation in PKS 0420$-$01 is relatively weak, but
still significant; however, no significant correlation is found
for FSRQ 3C 345. These anti-correlations imply that the spectrum
becomes redder (steeper) when the source is brighter.
%This tendency can be
%clearly seen from the upper and middle panel in Fig. 19.
%Particularly, the V-I color index of September are smaller (bluer)
%than those of other three months, in the case that object was
%faintest on September during our observation runs.
However, this result is opposite to the common colour change
tendency in blazars, that the spectrum becomes flatter when the
source brightens (Ghisellini et al. 1997; Ghosh et al. 2000;
Clements \& Carini 2001; Raiteri et al. 2001; Villata et al.
2002). In contrast to our FSRQs, we find that all our BL Lac
objects tend to be bluer (flatter) when they are brighter. We find
significant positive correlations between the colour indices and
$R$ magnitudes for 3C 66A, S5 0716+71, and BL Lac, with linear
Pearson correlation coefficients of $r=0.511$, $r=0.438$, and
$r=0.657$, respectively, at $\gg99.99$ per cent confidence; while
only weak correlations are found in PKS 0735+17 and OJ 287, at
very low confidence levels. The different trends of the colour
index variations with brightness in FSRQs and BL Lac objects are
somewhat unusual. However, this may indicate that there are
different physical conditions in these two populations, although
all of them were selected as being red blazars in the present
work. Nevertheless, it should be noted that our results are based
on limited data which span only a short period (about three
months). Certainly, more observations are needed to reinvestigate
the relationship between the colour index variations with the
source brightnesses for our sample. In particular, it would be
necessary to confirm and/or check the possible anti-correlation
for FSRQs with more observations, and then to further explore the
observed differences between FSRQs and BL Lac objects.

\begin{table*}
\caption{The correlations between $V-I$ colour index and $R$
magnitude, for the eight blazar objects.}
\begin{tabular}{lcccc}
\hline \hline
Source & $\Delta R$ & $N$ & $r$ & Probs.\\
\hline

3C 66A          &  0.50  &  120  &   0.511 &  7.34$\times10^{-10}$ \\
PKS 0420$-$01  &  1.48  &  55   &  -0.388 &  2.62$\times10^{-3}$  \\
S5 0716+71     &  1.37  &  271  &   0.438 &  1.11$\times10^{-14}$ \\
PKS 0735+17    &  0.54  &  38   &   0.111 &  4.97$\times10^{-1}$  \\
OJ 287          &  1.40  &  29   &   0.354 &  4.99$\times10^{-2}$  \\
3C 345          &  0.55  &  22   &  -0.009 &  9.66$\times10^{-1}$  \\
BL Lac          &  1.05  &  53   &   0.657 &  1.33$\times10^{-8}$  \\
3C 454.3        &  0.93  &  48   &  -0.837 &  1.07$\times10^{-16}$ \\

\hline
\end{tabular}
\end{table*}

%The plots of the resulting V-I indexes as a function of time, as
%well as R magnitude with time, and V-I with R magnitude are shown
%in Fig. 6. One can see that the color index is indeed variable,
%but the spectral changes do not follow the long-term brightness
%variations.
As shown in Sect. 3.3, variations on different time scales have
been found for the most extensively monitored source in our
campaign: S5 0716+71. We then explored the correlation between the
colour indices and the source brightness on different time scales.
It can be seen from Table 4 that there is a significant
correlation with linear Pearson correlation coefficient $r=0.438$,
using the whole data set, which spans about 140 days. During the
rapid brightening over the 4 days of 19 - 23 November 2003 (Fig.
4), we find a significant correlation with linear Pearson
correlation coefficient $r=0.656$, which suggests that the trend
of flatter spectra when brighter also holds for variations on time
scales of days. This spectral flattening with increasing
brightness was also recognized by Villata et al. (2000) in the
72-hour optical light curves obtained for S5 0716+71, during the
WEBT campaign of February 1999. However, we do not find any
significant correlations between the colour index and brightness
during the intraday variations, either on 23 November 2003 (Fig.
5) or on 10 December 2003 (Fig. 6). Ghisellini et al. (1997)
detected a spectral flattening when the flux is higher during
rapid flares; however, no correlation between spectral index and
brightness level was found in the long-term trend. A curved
trajectory of a relativistic emitting blob, or very rapid electron
injection and cooling processes, are proposed by the authors to
interpret the fast variations. Recently, using a large database,
Raiteri et al. (2003) found that the optical colour indices are
only weakly correlated with brightness, and that a clear spectral
steepening trend was observed during at least one long-lasting
dimming phase (JD $\sim$ 2451545 - 2451630). However, different
spectral behaviours were found on shorter time scales.
%Moreover,
%the optical spectrum became steeper after JD $\sim$ 2451000, the
%change occurring in the decaying phase of the late-1997 outburst.

In the case of 3C 66A, the trend of flatter spectra when brighter
is consistent with the recent results of Vagnetti et al. (2003),
who have found a consistent trend of $B-R$ hardening with
increasing $B$ band flux, independent of the actual flux value.
Ghosh et al. (2000) found that the spectral index of the source,
between $V$ and $R$ bands, flattened when the source brightness in
$V$ remained almost constant, while decreasing in $R$, between
1997 December 15 and 1998 January 2. Also, between 1998 January 19
and 30, the source brightness decreased, while the $V-R$ colour
index reddened, and that this correlated well with the variation
of the source brightness. However, from an analysis of a much
larger data set of an extensive multiwavelength WEBT monitoring
campaign from July 2003 through April 2004, B\"{o}ttcher et al.
(2005) found that there is a weak indication of a positive
hardness-intensity correlation at low flux states with $R\ga14.0$,
whereas no correlation is apparent at higher flux levels. These
authors claimed that this might be a consequence of the fact that
the $B-R$ hardness actually peaks during the rising phase of
individual outbursts.

\section{Discussion}

%Strong scattering of V-I colour index of PKS 0420-014 is evident.

It is notable that our FSRQs in general, and 3C 454.3 in
particular, became redder when brighter (see Fig. 12). In other
words, the spectrum became steeper when the object was brighter,
and flatter when fainter. This behaviour is opposite to the common
trends for blazars, that they become bluer when they brighten
(Ghisellini et al. 1997; Fan et al. 1998; Massaro et al. 1998; Fan
\& Lin 1999; Ghosh et al. 2000; Clements \& Carini 2001; Raiteri
et al. 2001; Villata et al. 2002). It has been found by different
investigators that the amplitude of the variations is
systematically larger at higher frequencies, which suggests that
the spectrum becomes steeper when the source brightness decreases,
and flatter when it increases (Racine 1970; Gear, Robson \& Brown
1986; Ghisellini et al. 1997; Maesano et al. 1997; Massaro et al.
1998). Recent investigations on spectral variability also show
this general trend (D'Amicis et al. 2002; Vagnetti et al. 2003;
Fiorucci, Ciprini \& Tosti 2004). From our investigations, the
variations of the colour indices of our BL Lac objects indeed
follow this trend, e.g. 3C 66A, S5 0716+71, and BL Lac. This
common phenomenon may be explained in different ways (Fiorucci et
al. 2004). It may indicate the presence of two components that
contribute to the overall emission in the optical region, one
variable (with a flatter slope $\rm \alpha_{var}$, where
$f_{\nu}\propto\nu^{-\alpha}$), and the other stable (with $\rm
\alpha_{const}>\alpha_{var}$). It is also possible to explain it
with a one-component synchrotron model: the more intense the
energy release, the higher the particle's energy (Fiorucci et al.
2004). Moreover, it could also be explained if the luminosity
increase was due to the injection of fresh electrons, with an
energy distribution harder than that of the previous, partially
cooled ones (e.g. Kirk, Rieger \& Mastichiadis 1998; Mastichiadis
\& Kirk 2002). In addition, it could be due to a Doppler factor
variations on a spectrum slightly deviating from a power law, e.g.
Doppler factor variations on a `convex' spectrum (Villata et al.
2004). It may also be possible that more than one mechanism are at
work, as pointed out by Villata et al. (2004), the variability
observed in the optical curves of BL Lac can be interpreted in
terms of two components: a `mild-chromatic' longer-term component
and a `strongly-chromatic' shorter-term one, which can be likely
due to Doppler factor variations on a `convex' spectrum and
intrinsic phenomena, such as particle acceleration from
shock-in-jet events (e.g. Mastichiadis \& Kirk 2002),
respectively.

However, some evidence that the amplitudes of variations are not
systematically larger at high frequencies has been found on
several occasions (see, for example: Malkan \& Moore 1986; Brown
et al. 1989; Massaro et al. 1998; Ghosh et al. 2000; Clements et
al. 2003; Ram\'{i}rez et al. 2004). Based on their results, Ghosh
et al. (2000) suggested that it may not be correct to generalize
that the amplitude of the variation in blazars is systematically
larger at higher frequency. In particular, they found a reddening
(i.e. spectral steepening) in their optical observations of the BL
Lac object PKS 0735+17. However, our analysis of this source shows
that there is a tendency of blueing (i.e. spectral flattening) as
the brightness increases, although the correlation between the
colour indices and brightnesses is rather weak (see Fig. 12 and
Table 4). Moreover, Ghosh et al. (2000) found that the spectral
slope of AO 0235+164 remained almost constant when its brightness
increased. They concluded that these characteristics cannot be
described simply by energy losses in a pure synchrotron mechanism
scenario.

The trend that FSRQ PKS 0736+017 was redder when brighter was
noted by Clements et al. (2003). They concluded that the
phenomenon appeared to be more related to the nature of the
variation, than to the host galaxy or the source brightness.
Recently, Ram\'{i}rez et al. (2004) detected the same peculiar
tendency to redden with increased brightness in PKS 0736+017,
throughout their observations. In addition, the analysis of their
data, and of the data reported by Clements et al. (2003),
suggested two varying modes. At low flux levels, small changes in
flux correspond to large changes in the spectral slope, while much
less pronounced spectral changes correspond to a high brightness
state. In both cases, the object reddens when it brightens.

FSRQs usually show strong emission lines, and a thermal
contribution that may be comparable to the synchrotron emission in
the optical spectral region. Their optical emission is
contaminated by thermal emission from the accretion disk and the
surrounding regions. Of particular relevance is the presence of
the so-called ``blue bump", or ``UV bump", which flattens the
spectral slope in the optical region. Since the thermal
contribution is larger in the blue region, the composite spectrum
would be flatter than the non-thermal component. Then, when the
object is brightening, the non-thermal component has a more
dominant contribution to the total flux, and the composite
spectrum steepens. This scenario has been used to qualitatively
explain the low flux level data of PKS 0736+017 (Ram\'{i}rez et
al. 2004). It can also be applied to explain the general trends of
steeper spectra when brighter in our FSRQs.

Contrary to the large majority of AGNs, which are characterized by
optical spectra with prominent emission lines, BL Lac objects have
quasi-featureless spectra. In fact, by the definition of this
class of AGNs, the line equivalent widths should be very small. As
originally proposed by Blandford \& Rees (1978), the weakness of
the spectral lines is most probably due to the fact that the
underlying non-thermal continuum is boosted by the relativistic
beaming of a jet pointing in the observer's direction. In
addition, the central ionizing luminosity of BL Lac objects may be
relatively too weak to produce prominent emission lines.
Consequently, the thermal contribution is rather small in the
optical spectral region, as compared to the synchrotron emission,
even when the source is in a low state. In actuality, Vagnetti et
al. (2003) found that BL Lac objects and quasars are clearly
segregated in the $\alpha$-$\beta$ plane (see their Fig. 3), where
$\alpha$ is the average spectral slope, and $\beta$ is a spectral
variability index. The lower average spectral index of their eight
BL Lac objects indicates the absence, or lower relative weight, of
the thermal blue bump component. The authors further proposed that
the segregation in the $\alpha$-$\beta$ plane is a consequence of
the different emission mechanisms in the optical band: synchrotron
in the case of BL Lac objects, and thermal hot spots on the
accretion disk in the case of QSOs. They found that a simple model
representing the variability of a synchrotron component could
explain the spectral changes of their BL Lac objects. Our BL Lac
objects were selected to be Low-energy-peaked BL Lacs (LBLs);
thus, we expect to observe them in the descending part of the
spectral power distribution, blueward of the peak frequency. The
strong variability of BL Lac objects is generally attributed to
the synchrotron emission. It is thus not surprising that they
generally follow the common trend for blazars, i.e. bluer when
brighter, as we observed for our BL Lac objects.

Although the thermal contribution may in general be relatively
large in the optical spectral region in FSRQs, the amount of this
contribution can be quite different from source to source. It is
conceivable that the thermal contribution is larger in 3C 454.3,
as compared to 3C 345, in which only a weak correlation is found.
It might be interesting to estimate the thermal contribution in
these sources, to better understand the tendency of the colour
index to vary with brightness. Nevertheless, it is not easy to
distinguish the thermal from the non-thermal components using
optical data alone.
%Also, the host galaxy contribution should be excluded in
%calculating the colour index and/or spectral slope, because the
%thermal contribution of the host galaxy, if it is not negligible,
%may produce a steepening of the energy distribution in the optical
%region (see, e.g., Pian et al. 1994).
It is essential to obtain simultaneous observations completely
covering at least the near-IR-to-optical part of the spectrum, to
better understand the spectral variability of FSRQs and/or
blazars. There is no doubt that more observations of blazars in
general, and those blazars that show reddening as they brighten in
particular, are needed to investigate the relationship between
colour index and/or spectral slope and brightness in more detail.

\section{Summary}

We present the results of our monitoring of the flux variability
of eight red blazars from 2003 September to 2004 February. The
main results can be summarized as follows:

1. All sources showed strong variabilities with amplitudes $>0.5$
mag during our 3-4 month monitoring period. Variations with
amplitudes of over 1 mag are found in four sources. In the extreme
cases, i.e. PKS 0420$-$01 and OJ 287, large variations with
amplitudes of over 1.4 mag were observed.

2. Intraday variations were found in S5 0716+71. Variations with
amplitudes $>0.15$ mag were detected on November 23 and December
10, 2003. A rapid rising rate of 0.002 mag per min was found on
December 10, 2003. We found a mean behaviour of a rapid linear
brightening with a rising rate of $\sim0.2$ mag per day over 4
days, during November 19-23, 2003.

3. We found that two out of three FSRQs tend to be redder when the
sources are brighter; and, conversely, all our BL Lac objects tend
to be bluer. There is a significant anti-correlation between the
$V-R$ colour index and $R$ magnitude for 3C 454.3, which is
opposite to the common colour change tendency in blazars. For our
3 BL Lac objects, 3C 66A, S5 0716+71, and BL Lac, there are strong
positive correlations, while only very weak correlations were
found for the remaining 2 BL Lac objects, PKS 0735+17, and OJ 287.

4. We propose that the different relative contributions of the
thermal versus non-thermal radiation to the optical emission may
be responsible for the different trends of the colour index with
brightness in FSRQs and BL Lac objects.

%We propose that the different contribution of thermal emission
%in the optical region may be responsible for the different trends
%of the spectral with brightness in FSRQs and BL Lac objects.

\begin{acknowledgements}

We are thankful to Dr. Jinming Bai (Yunnan Observatory, Chinese
Academy of Sciences, China) for stimulating discussions that
helped to improve this paper. We thank the anonymous referee for
insightful comments and constructive suggestions. We owe great
thanks to the observatory staff who supported our blazar
monitoring campaign: Bohyunsan Optical Astronomy Observatory,
Sobaeksan Optical Astronomy Observatory, and Mt. Lemmon
Observatory. This work was partly supported by the National
Natural Science Foundation of China (grants 10103003, 10373019 and
10543002) and the Korea Science and Engineering Foundation (grant
F01-2005-000-10209-0).

\end{acknowledgements}

\end{document}